%% file: palmer_elroi_aiaa_2017.tex
\documentclass{AIAA}

\usepackage{graphicx} 
\usepackage{subfigure} 
\usepackage{amsmath} 
\usepackage{siunitx} 
\usepackage{textcomp} 

\DeclareSIUnit \dbg {dB\gamma}
\DeclareSIUnit \dbm {dBm}

\begin{document}

\title{ELROI: A License Plate For Your Satellite}

\author{David M. Palmer\footnote{ISR-2, MS B244, Los Alamos National Laboratory, \texttt{palmer@lanl.gov}} and Rebecca M. Holmes\footnote{ISR-1, MS D466, Los Alamos National Laboratory, \texttt{rmholmes@lanl.gov}}}
\affiliation{Los Alamos National Laboratory, Los Alamos, NM, 87545}


\begin{abstract}
Space object identification is vital for operating spacecraft, space traffic control, and space situational awareness, but initial determination, maintenance, and recovery of identity are all difficult, expensive, and error-prone, especially for small objects like CubeSats. Attaching a beacon or license plate with a unique identification number to a space object before launch would greatly simplify the task, but radio beacons are power-hungry and can cause interference. This paper describes a new concept for a satellite license plate, the Extremely Low Resource Optical Identifier or ELROI. ELROI is a milliwatt-scale self-powered autonomous optical beacon that can be attached to any space object to transmit a persistent identification signal to ground stations. A system appropriate for a LEO CubeSat or other small space object can fit in a package with the area of a postage stamp and a few millimeters thick, and requires no power, data, or control from the host object. The concept has been validated with ground tests, and the first flight test unit is scheduled for launch in 2018. The unique identification number of a LEO satellite can be determined unambiguously in a single orbital pass over a low-cost ground station.
\end{abstract}

\maketitle

\section*{Nomenclature}
\noindent\begin{tabular}{@{}lcl@{}}
\textit{$\lambda$} &=& wavelength\\
\textit{$\Delta\lambda$} &=& spectral bandwidth of filter\\
\textit{$\tau$} &=& pulse duration\\
\textit{$T$} &=& pulse interval\\
\textit{$P_{peak}$} &=& peak emitted optical power\\
\textit{$P_{ave}$} &=& average emitted optical power\\
\textit{$\varepsilon_{\textrm{DQE}}$}&=& detector quantum efficiency, the ratio of detected to incident photons\\
\textit{$R_{dark}$}&=&detector dark rate\\
\textit{$\Omega$} &=& emission solid angle\\
\textit{$D$} &=& telescope effective aperture diameter\\
\textit{r} &=& range from satellite to ground station\\
\si{\dbm} &=& decibels referenced to \SI{1}{\milli\watt}\\
\si{\dbg} &=& decibels referenced to \num{1} \si{photon\per\s}\\
\end{tabular} \\

\section*{Terminology and Acronyms}
\noindent\begin{tabular}{@{}lcl@{}}
\textit{ELROI} &:& Extremely Low Resource Optical Identifier\\
\textit{\$SWaP}  &:& Cost, Size, Weight, and Power \\
\textit{HRT} &:& Horizontal Range Test\\
\textit{ID} &:& Satellite Identification Number\\
\textit{LEO} &:& Low Earth Orbit\\
\textit{RFI} &:& Radio Frequency Interference\\
\textit{STC} &:& Space Traffic Control\\
\textit{SOI} &:& Space Object Identification\\
\textit{SSA} &:& Space Situational Awareness\\
\end{tabular} \\

\section{Introduction}

Space is crowded---with thousands of manned spacecraft, active and retired payloads, rocket bodies, explosion fragments, and other orbiting space objects---and getting more so. Knowing where satellites are is essential for preventing collisions, and useful for satellite owners and operators. The current publicly available Space-Track catalog \cite{spacetrack} lists over 16,000 objects along with their orbital elements.  However, this list is restricted to those objects that have been continuously tracked since launch to maintain ``track custody,'' or those that have been confidently identified by other means.  Track custody can be lost when there are interruptions in observations, when solar activity causes sudden increases in atmospheric drag, when there are unexpected maneuvers, or when two satellite orbits become confusingly close.  When an unidentified satellite then reappears in radar and optical observations, it can't be put back into the catalog unless multiple observations can quantify its orbit and track it back to a previously-lost satellite. Even when a satellite is tracked from the moment it is released from its launcher, it may not be fully identified.  Individual rocket launches can release dozens to more than a hundred satellites (particularly CubeSats) at a time \cite{Foust2017a,Foust2017}.  Shortly after launch, a satellite operator may know only that their satellite is one of the many ``OBJECT XX'' entries in the catalog, making it difficult to contact and control the spacecraft.

Space Object Identification (SOI) is easier if a satellite carries an identifying beacon that can be read from the ground. Ideally, this beacon would consume few resources and would last the entire orbital lifetime of the satellite. There is currently no standard or widely accepted technology to provide such a beacon, but several options are under consideration, including conventional radio beacons combined with GPS, radio tags activated by a ground signal, and passive reflectors \cite{Ewart2016}. Radio is the most developed option, but radio transmitters that can be received at orbital distances tend to be heavy and power-hungry. Furthermore, if they operate continuously they can cause radio frequency interference (RFI), so they are turned off when a satellite is decommissioned. Thus, radio beacons do not enable SOI for post-mission satellites, and they are also not suitable for passive debris objects such as rocket boosters and interstages. Conventional optical beacons (LEDs, for example) avoid the problem of RFI, but still require significant power to be visible from the ground with traditional detection methods.

\begin{figure}[p!]
\includegraphics[width=0.75\textwidth]{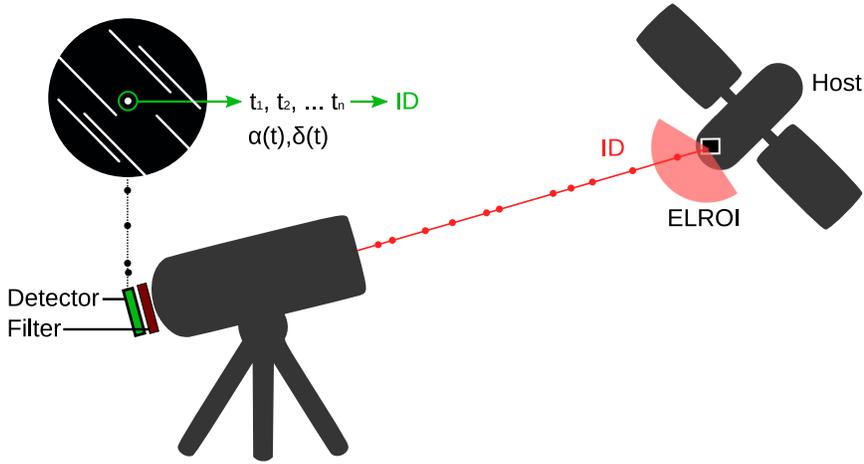}
\caption{Overview of the ELROI system. The beacon is attached to a satellite and continuously emits its optical signal---encoding a unique ID number---over a wide solid angle. A ground telescope collects a small portion of the emitted photons, which are detected with a photon-counting sensor. A narrowband filter centered on the beacon wavelength rejects background light. The recorded data (circular inset) consists of a list of photon detection times at a tracked location (green circle). Streaks represent background stars. The data analysis technique uses the timing characteristics of the ELROI signal to eliminate more than 99\% of background photons, making it possible to read the ID in a single pass even if the signal is only a few photons per second.}
\label{figoverview}
\end{figure}

\begin{figure}[p!]
\includegraphics[width=0.75\textwidth]{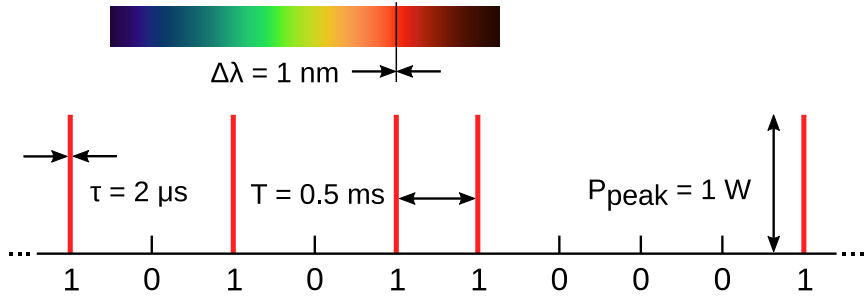}
\caption{Representative characteristics of the ELROI signal. The spectral bandwidth of the signal is about 1~nm, and the example wavelength shown is 638~nm (other wavelengths may also be suitable, as discussed in \S\ref{sect:design}). The width of the optical pulses is $\tau = 2\mu$s, and the interval between pulses is $T = 0.5$ ms. If the instantaneous peak power is 1 W and half the bits contain a pulse, the average power with this duty cycle is 2 mW, and a 128-bit ID repeats every 64 ms.}
\label{figwaveform}
\end{figure}

The Extremely Low Resource Optical Identifier (ELROI) beacon is a new concept for a low-power optical ``license plate'' that can be attached to anything that goes into space \cite{Palmer2015,Palmer2016} (Figures \ref{figoverview} \& \ref{figwaveform}). The ELROI beacon encodes a unique ID number in short, omnidirectional flashes of laser light, which can be read from the ground with only a few photons per second using single-photon detection and an innovative background-rejection technique. The remarkably low power level required for photon-counting optical communications has already been noted as a possibility for low power communication, even at interstellar distances \cite{Howard2000}. This strategy allows the beacon to be low cost, small, lightweight, and low power, so it fits the \$SWaP (Cost, Size, Weight, and Power) budget, even at the CubeSat level. The power requirements are low enough that the beacon can be self-powered with a small solar cell, which makes system integration simple, and allows the beacon to be flown on inert debris objects without power, communications, data, or attitude control systems. The beacon does not produce RFI, and may be operated continuously throughout the on-orbit lifetime of the satellite.

The ELROI signal consists of short, bright pulses of monochromatic light clocked at a fixed period. Each period starts with either a flash of light (encoding a 1 bit), or no flash (a 0), as shown in Figure \ref{figwaveform}. The ID repeats several times a second. The peak power of the light source (a laser diode) is in the 1 W range, but the average power is much lower. If the pulse duration and interval are as shown in Figure \ref{figwaveform} and half the bits are ones, the duty cycle of the laser diode is 1:500, and a peak power of 1 W gives an average power of just 2 mW. This means that power for the beacon can be provided by a few square centimeters of solar cell, making it independent of the host satellite.

The ELROI signal is intentionally open and accessible to anyone with a ground station that can be built from Commercial Off The Shelf components.  The identification number (ID) of each beacon will be stored in an open registry, along with contact information for its operator and other information.  This allows ELROI to be adopted as an international standard, read by ground stations around the world to assist in the worldwide problem of space traffic control (STC) and space situational awareness (SSA). The beacon can transmit additional data beyond the ID, giving satellite operators a backup channel for anomaly resolution and other diagnostic purposes. This, along with the benefits to the spacecraft operator of being able to identify their own satellite, can drive adoption of ELROI even in the absence of international norms or mandates. 

\begin{figure}[p!]
\subfigure[NMTSat]{\label{fig:a}\includegraphics[width=0.3\textwidth]{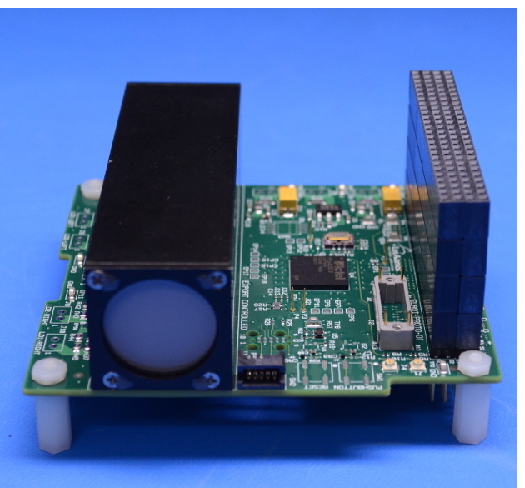}}
\subfigure[ELROI-UP]{\label{fig:b}\includegraphics[width=0.3\textwidth]{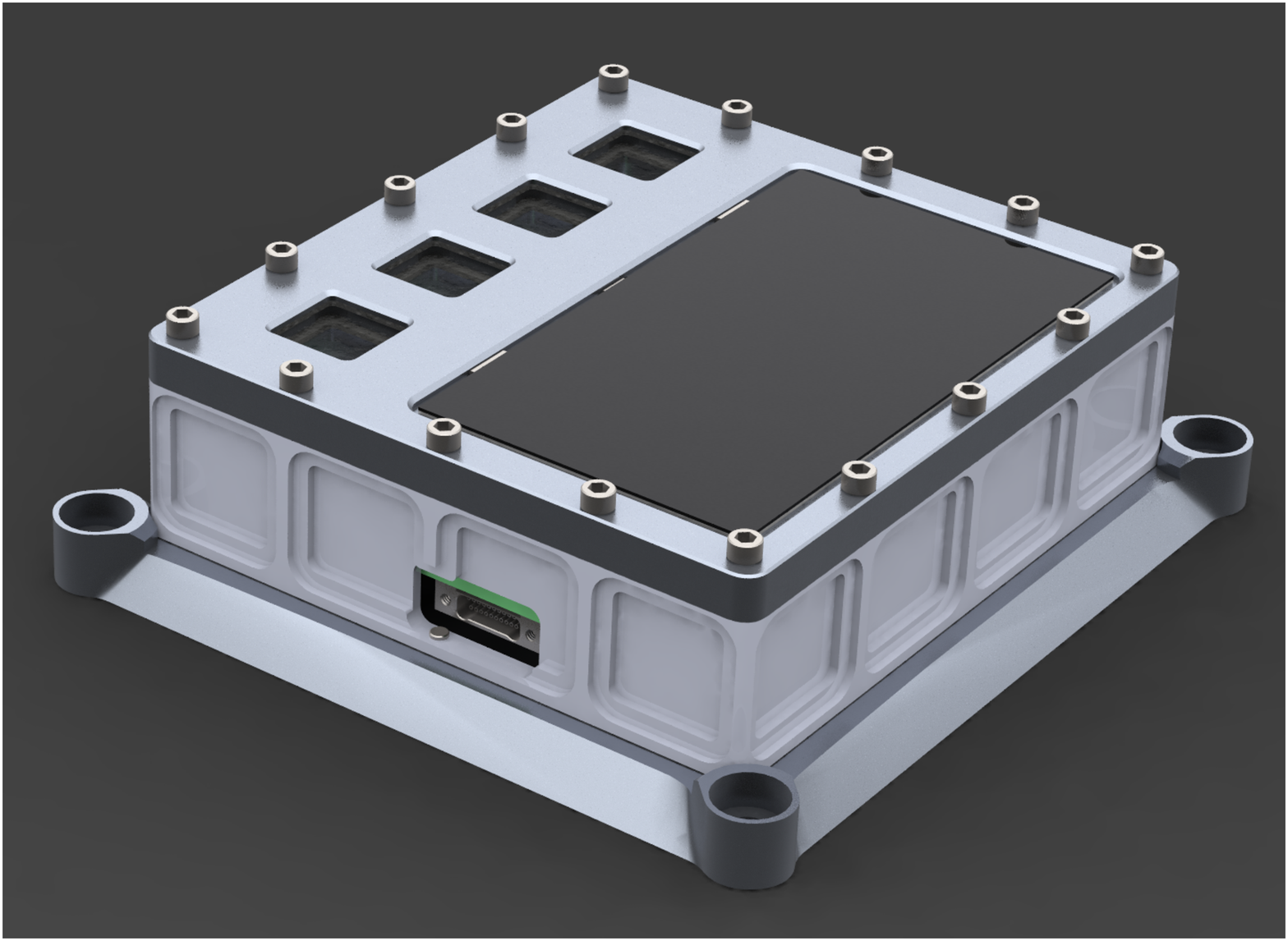}}
\subfigure[ELROI 1.0]{\label{fig:c}\includegraphics[width=0.3\textwidth]{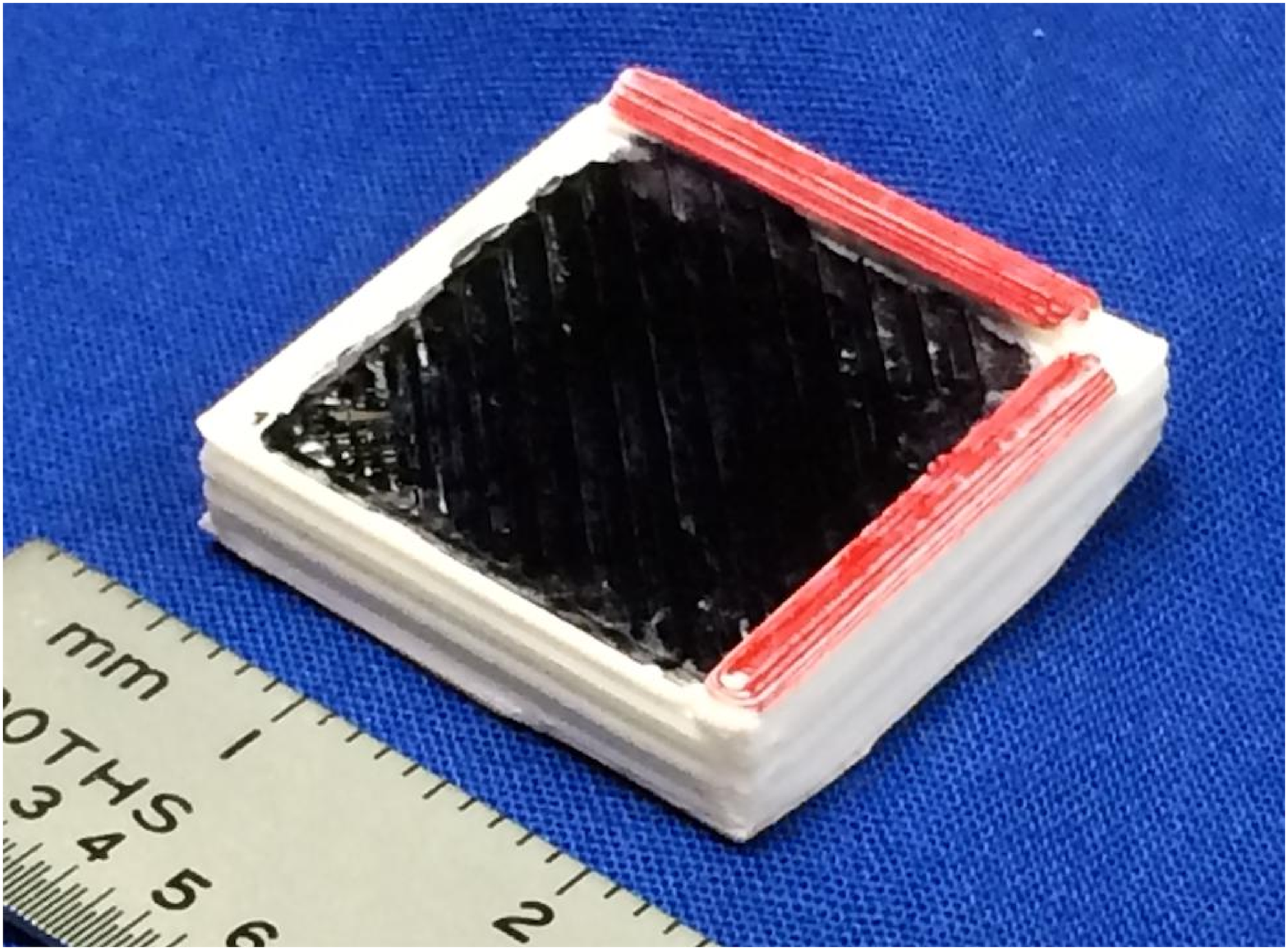}}
\caption{ELROI hardware development stages.
    {\bf a.} ELROI-PC104 unit to be incorporated into a CubeSat in the PC-104 form factor.  
    {\bf b.} CAD Drawing of ELROI-UP unit that can operate fully autonomously when attached to host object, or as a free-flyer.
    {\bf c.} Illustrative mock-up of a miniaturized ELROI unit, suitable for any LEO CubeSat or other small LEO space object.}
\label{fig:prototypes}
\end{figure}

The first ELROI prototype has been delivered and integrated with a spacecraft (Figure \ref{fig:prototypes}a) scheduled for launch in 2018.  A fully autonomous module (Figure \ref{fig:prototypes}b) suitable for either attaching to a host satellite or launchable as a 1/2U CubeSat free-flyer, is currently being designed and will be produced in quantity (6 for the initial production run) to support multiple flight opportunities as they present themselves. An optimized beacon will eventually be miniaturized to a package the size of a thick postage stamp---a few square centimeters in area and a few millimeters thick---that can be mass produced in quantity to support the global launch rate (Figure \ref{fig:prototypes}c).

The ELROI concept has also been verified in ground-to-ground tests. Using two test units operating at 638 nm with average power less than 0.3 \si{\milli\watt} at a distance of 15 km, photon detection rates in the range of 0.1--40 \si{photons\per\second} were measured, depending on lens aperture and attenuation.  As discussed in detail in \S \ref{hrt}, these measurements are consistent with the predicted detection rate, and the same model predicts a rate of approximately 3.3 signal \si{photons\per\second} from a LEO range of 1000 km (Appendix B). The encoded ID was successfully extracted from the ground-to-ground test data, with reliability depending on the total number of photons. Scaling the measured photon detection rate to simulate the LEO case shows that approximately 100-160 seconds of observation at the LEO rate is sufficient to reduce the misidentification rate to less than 1 in a billion, making it practical to read the ID in a single satellite pass over the ground station.

ELROI has been developed at Los Alamos National Laboratory (LANL), growing out of our experience both with space systems and with single-photon imaging. Initial testing will use a pre-existing ground station consisting of a commercial telescope (a Celestron 35 cm aperture telescope on a Software Bisque Paramount) and a LANL-developed single-photon imaging camera \cite{Priedhorsky2005,Thompson2013}.  However, any additional groups that wish to use their own systems to identify the ELROI beacons are invited and encouraged to do so.

In this paper, \S \ref{sect:concept} describes the ELROI concept, and briefly describes the design and operations of the beacon and the ground station.  \S \ref{sect:design} discusses some of the design details that allow a low power signal to be transmitted, what trade-offs and optimizations are available to make this a practical system, and what else can be done with ELROI.  Our progress so far in developing this, with both ground tests and current and future flight hardware, is discussed in \S \ref{sect:status}.  The Conclusion, \S \ref{sect:conclusion}, summarizes the paper and discusses how ELROI can develop to become a global standard.  Appendix A shows how to computationally extract the ELROI ID from the ground station's data and Appendix B works through an example link budget to demonstrate the feasibility of this concept.

\section{Concept}
\label{sect:concept}

ELROI uses a small optical beacon that can be attached to any object that goes into space (Figure \ref{figoverview}). This beacon produces flashes of light that encode an ID number, providing a ``license plate'' that uniquely identifies the satellite. The flashing light can be detected and read by a small ground telescope with a single-photon detector, allowing anyone to identify the satellite. The specific characteristics of the light flashes allow the ID to be read even if only a few photons are detected per second, and in the presence of background.

The optical signal is diffused to a large solid angle (almost a full hemisphere) so that spacecraft attitude control or pointing is not required.
The receiver telescope tracks the satellite, using its known orbital elements.  The light is spectrally filtered to select only those photons at the beacon wavelength.  A single-photon detector records the arrival time of each photon, and the list of times is then analyzed to reconstruct the ID by determining which time bins contain 1's and 0's.  This data analysis is described in more detail in Appendix A. 

The combination of the beacon signal characteristics and the ground station design enable background rejection and signal extraction techniques that allow an extremely low-power signal to be recovered.  Unlike a radio antenna, an optical telescope's imaging capability can reject all background sources more than a small fraction of a degree from the beacon.  The spectral purity of the laser diodes allow narrow-bandwidth filters to reject almost all of the non-beacon light from the sky and the satellite itself.  The use of single-photon detection and timing brings the signal into the digital domain, eliminating analog noise and allowing signal integration over the entire observation time.  The extremely low duty cycle and strict periodicity of the beacon allows a phase cut that rejects the background light that doesn't precisely match the timing of the beacon.  Finally, the use of an error-correcting coding scheme for the ID makes it very unlikely that one ID will be mistaken for another, even when the values of some bits are uncertain.

\subsection{System Design Overview}

More detailed design considerations are discussed in \S \ref{sect:design}.

\begin{figure}[p!]
\centering
\includegraphics[width=0.6\textwidth]{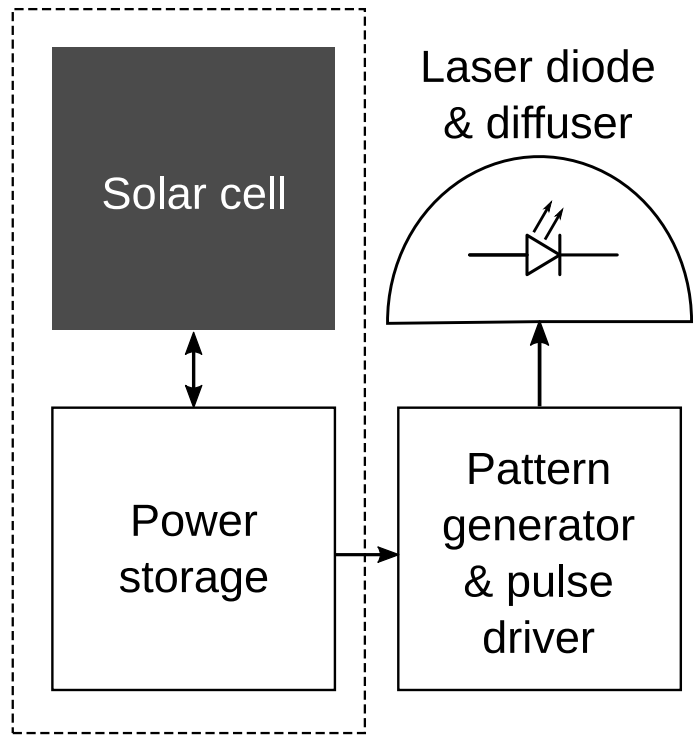}
\caption{Components required for the beacon. A small solar cell provides power independent of the host satellite, with the battery allowing operation during orbital night. A pattern generator supplies the pulse sequence to a pulse driver, which powers the laser diode that emits the optical signal.  The light is diffused into a wide solid angle so that the beacon does not need to be pointed directly at the ground station.
}
\label{figelectronics}
\end{figure}

The electronics required for the beacon are quite simple, as shown in the block diagram Figure \ref{figelectronics}. The power system is mostly remarkable for its low capacity requirements compared to conventional satellite systems. The pattern generator can be a very simple microcontroller, FPGA, or ASIC. The pulse driver circuitry drives a laser diode, which provides highly monochromatic light with good efficiency.  This light is diffused to cover a wide solid angle so that light reaches any ground station in its field of view without any pointing requirements placed on the host.

\begin{figure}[p!]
\centering
\includegraphics[width=\textwidth]{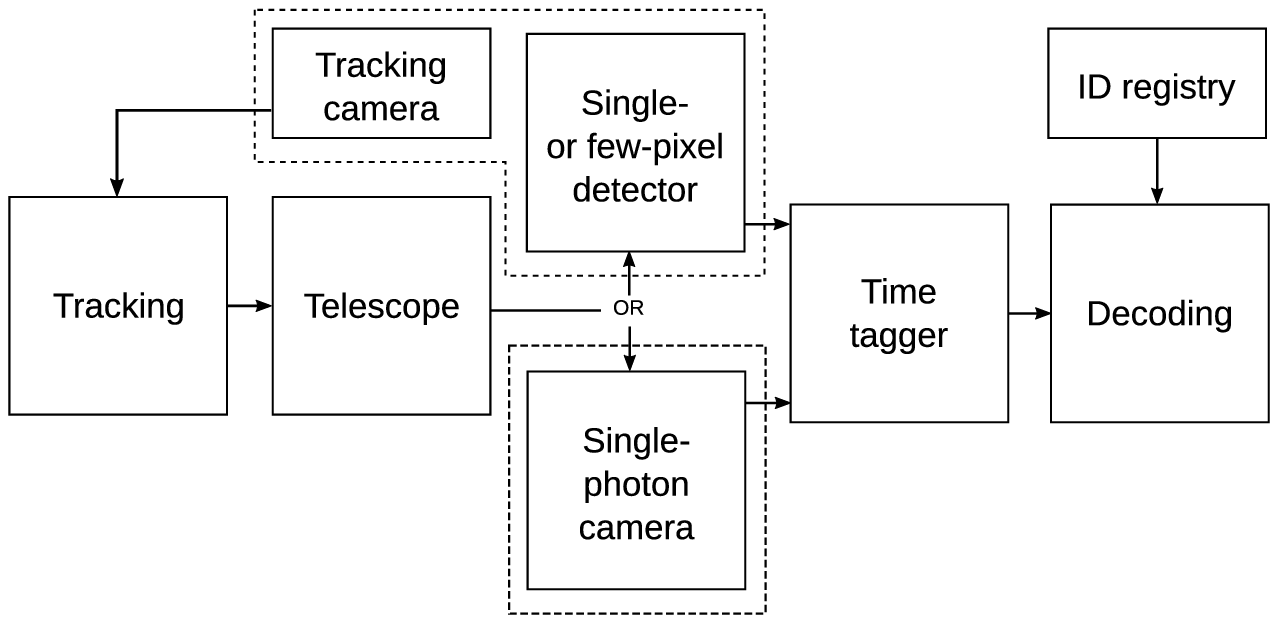}
\caption{Components required for the ground station. A tracking system drives a telescope equipped with a photon-counting sensor. If a single- or few-pixel detector is used, tracking requirements are more stringent than for an imaging sensor, and a separate tracking camera may be required. A time-tagger records the time of each photon detection. (For a single-photon camera, the location of each detection is also recorded.) The signal is decoded by acquiring the carrier frequency, rejecting out-of-phase background photons, and assigning a value to each bit based on the number of photons detected. The resulting bit values are compared to the entries in a registry to find the best match to an existing ID.}
\label{figgroundstation}
\end{figure}

The ground station, shown in Figure \ref{figgroundstation}, uses a telescope to concentrate the light from the satellite and reject light from other parts of the sky. This telescope must be mounted so that it can track satellites, based on supplied orbital elements. The collected light is further filtered by wavelength, using a narrow-band spectral filter at the beacon wavelength. This rejects the majority of  background light at other wavelengths (from the sky and from sunlight reflecting from the satellite).

The filtered light is then recorded with a ``photon-counting detector.''  This is a detector that, minimally, emits a pulse that triggers a time-recording circuit every time it detects a photon.  The detector may also have some position-sensitivity or imaging capability which, as we shall see in \S\ref{subsect:mountdetectortradeoff}, affects the requirements on the telescope and its tracking system. The data from the photon-counting detector---a list of photon times, including both beacon photons and background photons---is then analyzed to determine the ID of the beacon, as described in Appendix A.

\subsection{Operations}

The ELROI beacon requires a targeted observation to read an ID. It is intended as an adjunct to persistent satellite tracking operations by radar and optical methods that maintain and update a self-consistent catalog. It will typically be used for objects where the identity is unknown or questionable, such as immediately after launch when multiple satellites must be matched to multiple object detections. The existence of an ELROI beacon on an object relaxes the demands on the persistent tracking systems, because the consequences of a lost track are diminished if the identity can be recovered.

Due to sky background, the typical ground station can only operate at night.  Observations will require clear skies, at least over a significant portion of the apparent path of the satellite, but atmospheric turbulence (seeing) is unlikely to have much of an effect on receiver sensitivity.  The narrow field of view of the telescope will require that the satellite's orbit be known in advance.  The required orbit prediction accuracy is determined by the field of view of the photon counting detector, and the presence or absence of sunlight on the satellite to allow closed-loop tracking.  For larger satellites, observations in eclipse will reduce the background and allow for faster identification, at the cost of requiring accurate ``blind'' pointing without seeing the satellite itself in reflected sunlight.

Each ground station is expected to observe many different satellites during their respective passes above the horizon.  The cost of an individual ground station (likely to be in the in the range of 0.1--1 million dollars), and the skills required to operate it, will be beyond the practical capability of most individual CubeSat operators.  It would be wasteful to develop a ground station and use it only for a single satellite (which might be visible from a given ground station for only a few minutes a day) and that restricted usage would require that the satellite's identity be already known.  Thus ELROI identification is probably best implemented to by specialized ground station operators contacting the operator of each satellite as it is identified, as either a commercial or governmental service.
 
A full implementation of ELROI for global Space Traffic Control (STC) will require a worldwide network of ground stations to provide prompt identifications for a variety of orbital geometries, regardless of the weather and other conditions at individual ground stations.  Different locations, or different ground stations at a single location, can have different design choices in terms of telescope, mount and detector, to cover the diversity of situations expected. 

\section{Design and Trade-offs}
\label{sect:design}

\subsection{Beacon Design}
\label{subsect:beacondesign}

The optimal ELROI beacon signal characteristics---wavelength, peak power, pulse width and separation, etc.---will depend on trade-offs between the resources that will be devoted to each beacon (\$SWaP), the resources for each ground system (primarily cost), the characteristics of the satellite, and the required reliability of identification. Some of these characteristics must be standard for all ELROI units, while others can vary from satellite to satellite.

The strictest technical requirements for an ELROI standard are merely that the beacon produce sufficient light at an agreed-upon wavelength, so that a telescope with a narrow-band filter and a photon-counting detector can record the photons.  Variation in the details of the timing and coding scheme can be drawn from a small set of standards, which can evolve with time as more experience is gained and more capabilities are required.  (The computational power required to process a list of photon time-tags is minimal, allowing the data to be tested against each of the standards in use to find the correct one.)

Semiconductor laser diodes at the \si{watt} scale are readily available in rugged packages for a variety of technologies, and are relatively resistant to radiation at the levels expected in a LEO environment \citep{Johnston2000,Phifer2004,Ott2006}. The planned flight tests (\S \ref{pc104}--\ref{elroi1}) include a variety of laser diodes and, along with additional environmental tests, will help determine final part selection. Because this application does not require coherence, multiple independent emitters can be combined to provide the desired optical power.  This allows $P_{peak}$ to be scaled to arbitrary levels by merely adding more emitters.

Increasing $P_{peak}$ while keeping $P_{ave}$ constant (a beacon with shorter, brighter pulses) will reduce the number of in-phase background counts while keeping the source counts the same.  Thus, for larger (brighter) satellites, higher $P_{peak}$ laser diodes may be optimal if the package size is determined by the solar cell and battery (and hence $P_{ave}$).  For smaller satellites, reduction of the already low in-phase background gives less advantage and may not be worth the increased circuitry required for higher $P_{peak}$. 

Different laser technologies will have different characteristics, including power efficiency, availability of components, robustness, radiation hardness, and wavelength variability.  They will also have different wavelengths, which affects both the propagation characteristics and the availability and efficiency of detectors.  (The choice of $\lambda =$ 638 nm and 450 nm laser diodes for the initial flight units is largely determined by the wavelength-dependent sensitivity of the pre-existing LANL ground station and camera.)

If the beacon is a single unit mounted on a flat surface, this restricts emission to a hemisphere ($\Omega  = 2\pi$ \si{\steradian}) which for a randomly-oriented craft leads to a 50\% probability of being observable from a ground station at any given time. This can be addressed in a number of ways: the emitter may be mounted on a corner of the satellite or at the end of a solar panel to provide a wider view; multiple emitters or multiple beacons can be mounted around the spacecraft; or the restricted emission can be accepted as a factor that reduces but does not eliminate the probability of a valid ID. If, for example, the satellite is expected to be freely tumbling so that its emitter is visible to the ground station for half of a typical pass, then the aspect and attitude variation of a satellite during a given pass may be enough to provide a valid ID acceptably often.

Although Fig. \ref{figelectronics} shows the components of an autonomous beacon, some of the functions can be integrated with other spacecraft systems.  
A system that taps a small amount of power from the existing spacecraft bus and is controlled by surplus capacity of a processor or FPGA could be incorporated into a spacecraft during the design stage, requiring nothing more than the laser diodes, drivers, and diffusers to be added to the spacecraft hardware.  However, an autonomous beacon as a separate package can be added to a spacecraft much later in the design cycle with a much lower engineering expense.

The design of an error-correcting coding scheme that selects the ID numbers is beyond the scope of this paper. One possibility is a Bose-Chaudhuri-Hocquenghem (BCH) code, which for a 127-bit ID length would provide 4 million different IDs while being robust to up to 21 erroneous bits.  Another possibility is a 32-of-128 code, which has less error correction capability but (due to the lower number of '1' bits) requires less power to transmit.

\subsection{Ground Station Design}
\label{subsect:mountdetectortradeoff}

Because clear, dark skies with a line of sight to the satellite are required for a ground station to read the ELROI beacon, a global network of ground stations will be required to support the need for timely identification.  It is likely that there will be diversity in the types of detectors and other characteristics of these ground systems. The major design trade-offs are among the capabilities and costs of the detector, the telescope, and its tracking system.

The sensitivity of the ELROI ID measurement relies on the use of a ``photon-counting detector''.  That is, each photon registered by the detector must produce a discrete signal whose timing can be measured with resolution better than the pulse width $\tau$.  The detector will have a quantum efficiency, $\varepsilon_{\textrm{DQE}}$, with which it registers the photons that hit it, and a detector dark count rate, $R_{dark}$, which is the rate of false photon detections.

These detectors can be single pixels, registering the time for a photon anywhere on a sensitive area, or position sensitive detectors that provide both the time and a location in the field of view for each photon.  Popular single-pixel detectors include single-photon avalanche diodes (SPADs) and photomultiplier tubes (PMTs).  Arrays of SPADs, position-sensitive readouts for PMTs, and other technologies such as microchannel plates (MCPs) can provide position resolution, from simple quadrant detectors up to full imaging detectors with hundreds or thousands of pixels on a side.  (More familiar imaging sensors such as conventional CCDs, electron-multiplying CCDs, and CMOS imagers do not have the combination of time resolution and low read noise needed for reading an ELROI beacon.)

A single-pixel detector requires that the light from the satellite be concentrated on the detector, without too much additional background light.  This requires a telescope and mount that can precisely follow the track of a satellite across the sky.  If the satellite is illuminated by the Sun, a conventional camera can detect the satellite and provide pointing corrections to the telescope mount, while a dichroic mirror directs photons at the beacon wavelength to a single-photon detector.  If the satellite is not illuminated, the same conventional camera can provide detections of stars as they pass through the field of view, and keep the mount pointing at the satellite's \emph{predicted} location, based on orbital elements provided by previous observations. Some suitable ground stations already exist with the addition of a small amount of extra hardware and software---for example, satellite laser ranging (SLR) stations have the equipment and expertise to track LEO satellites and observe them with single-pixel detectors. These stations are widely-spaced around the globe and used for geodesy \cite{Pearlman2002}.

A position-sensitive detector allows a less-accurate tracking system to keep the satellite in a larger field-of-view, with the satellite photons separated from the star and sky background in later processing stages.  However, such detectors are substantially more complex and expensive than single-pixel detectors.  The ground station that we will use for the initial testing uses a photon counting camera developed at LANL \cite{Priedhorsky2005}, but equivalent systems are available commercially \cite{photonis, roentdek}.

A larger ground station aperture and a higher $\varepsilon_{\textrm{DQE}}$ increase both source and background count rates.  These cause a corresponding decrease in the time required to ID the beacon.

The trade-offs discussed above may drive the system to different designs in different operating regimes; for example, LEO vs. GEO satellites.  For observing LEO satellites, which have apparent speeds of order one degree per second across the sky, the optimum may be a relatively small telescope on a rapidly slewing mount, with an imaging detector used to reduce the required precision of the high-speed motion.  For satellites at or near geosynchronous orbit, a larger telescope can be mounted on a slower mount and locked on to a slowly-moving target to feed a single-pixel detector.  The larger telescope aperture and the higher $\varepsilon_{\textrm{DQE}}$ available in single-pixel detectors helps to compensate for the inverse-square loss at the much longer range to GEO.

Changes in technology will also change optimal design choices.  Large imaging SPAD arrays are under active development (driven by the desire for flash-LIDAR systems for self-driving cars \cite{Hecht2017,McManamon2017}) and may make high-performance photon counting cameras cheap. Superconducting nanowire singlephoton detectors (SNSPDs) are becoming commercially available and may provide higher performance than other single-pixel detectors at an acceptable cost \cite{quantum-opus,photon-spot,id-quantique}.

\subsection{Additional capabilities}

The ID signal consists of low-duty-cycle periodic flashes confined to an in-phase time slice of width $\tau$ in each period $T$.  This permits additional signals to be interleaved as extra flashes out of phase with the ID bits.  The additional signal must have low enough spectral power at the period of the ID signal to avoid interference with the clock recovery, but this can be achieved by a careful choice of coding scheme.

One possible use for this auxiliary data channel is to provide ``black box'' information for diagnosing and recovering from spacecraft anomalies.  The operation of an autonomous ELROI unit will be unaffected by most failure modes of its host satellite, and so this can provide an independent low-bandwidth channel for small amounts of status information when other information is not available. The ELROI unit may have its own sensors, such as MEMS accelerometers, gyroscopes, and shock sensors. Even with no additional hardware, variations in the solar cell voltage can be analyzed to detect sudden changes in satellite attitude or spin state.

The cost in \$SWaP for implementing this optional capability is relatively low, and includes an increase in complexity of the pattern generator, any interfaces or internal sensors that are required for the status information, and an increase in overall power proportional to the additional light generated to transmit the additional bits.

\section{Current results and future development}
\label{sect:status}

\subsection{Horizontal range test}\label{hrt}

\begin{figure}[p!]
\centering
\includegraphics[width=0.75\textwidth]{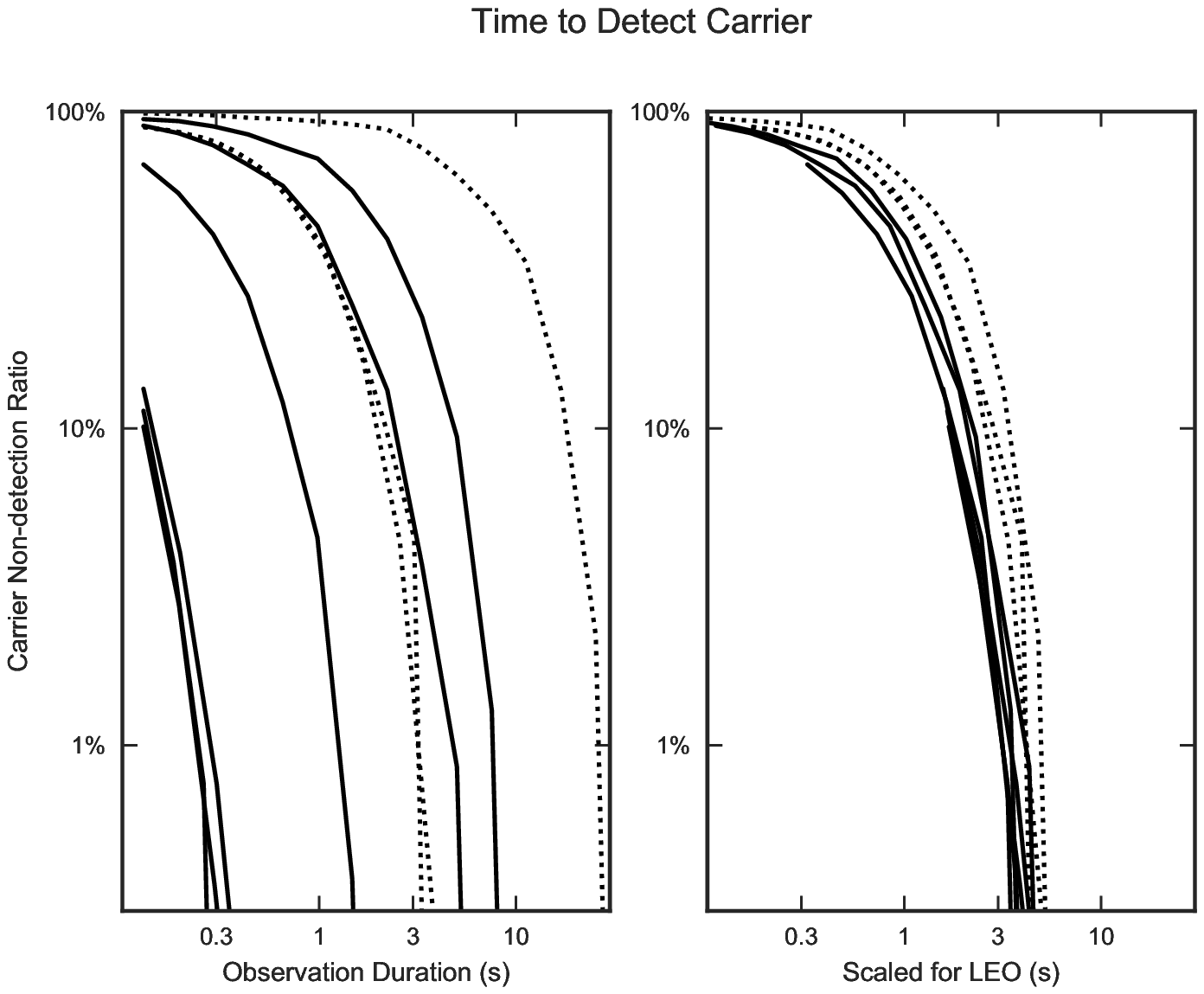}
\caption{Time required to detect the signal carrier period and phase. The left panel covers a range of apertures (f/2.8 to f/22) for both transmitter units (solid: Unit 1, dotted: Unit 2) in the HRT conditions, covering signal count rates of 0.1--40 counts/second.  The right panel is the same data, scaled to 3.3 signal counts/s to simulate the case of a LEO CubeSat.  The collapse of the scaled data to a common band gives us confidence in our link budget model and indicates that, after all the background rejection steps are implemented, the total number of signal counts is the dominant parameter in our ability to track the signal, at least in the CubeSat-equivalent case.}
\label{figcarrier}
\end{figure}

\begin{figure}[p!]
\centering
\includegraphics[width=0.75\textwidth]{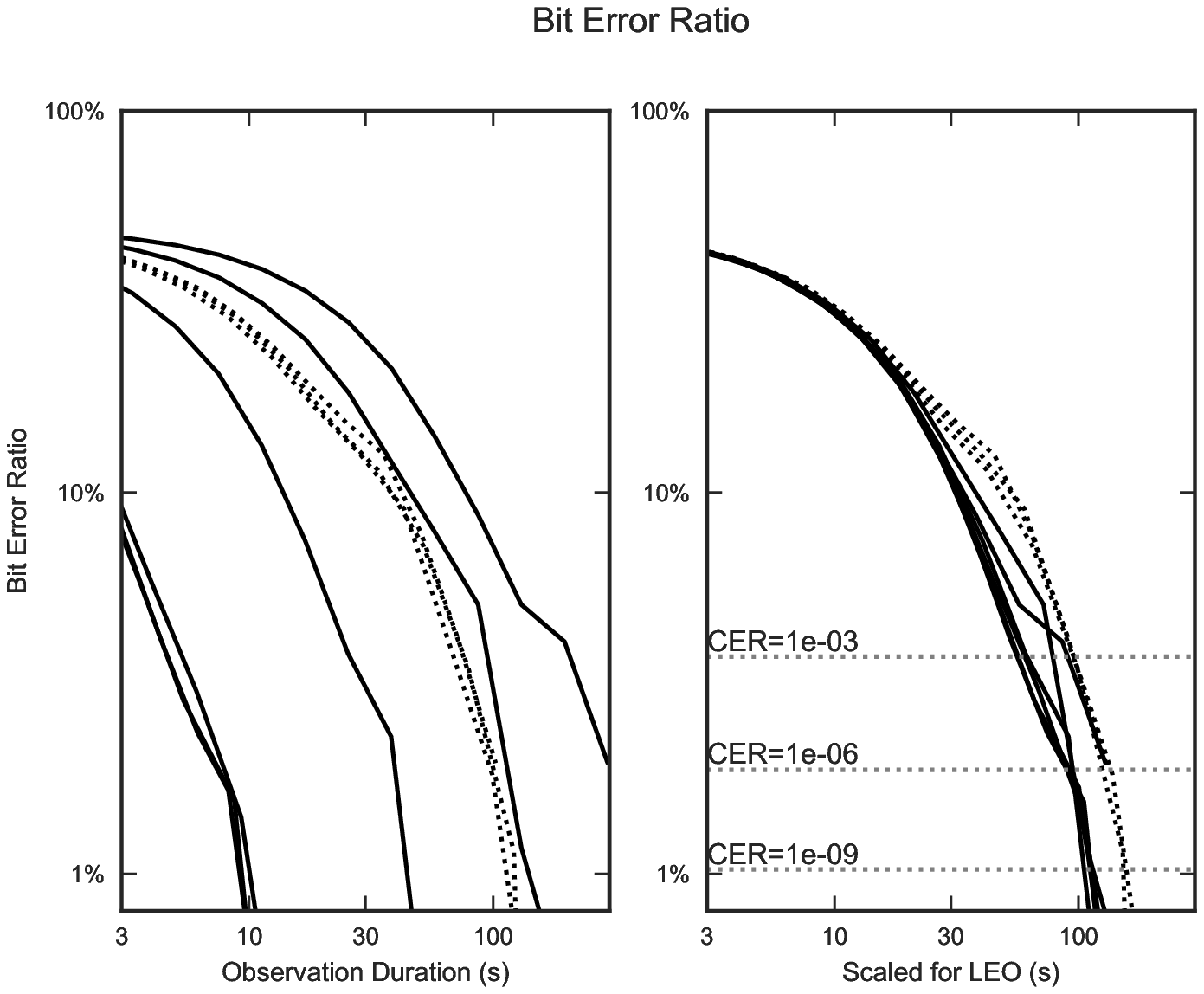}
\caption{Time required to achieve a confident ID read. The bit error ratio (BER) is shown as a function of observation duration in the HRT conditions (left), and scaled to simulate the LEO case (right).  The BERs corresponding to a range of codeword error ratios (CERs, horizontal lines) are marked.  As in Figure \ref{figcarrier}, the agreement among the scaled data measurements indicates that the dominant requirement for accurately reading the ID codeword is the number of signal counts.}
\label{figcoderror}
\end{figure}

To validate link budget calculations and determine whether the ID can be read in conditions similar to a LEO system, the ELROI concept was demonstrated with a horizontal range test (HRT) over approximately 15 km. The bit error ratio (BER) and resulting codeword error ratio (CER, the probability of incorrectly identifying the ID) were measured under a range of observing and analysis conditions to simulate different observation durations and signal strengths. The results of the HRT show that the link budget calculations are accurate when applied to the test conditions, and that the transmitted ID number can be read with data rates comparable to a LEO system.

\subsubsection{Horizontal range test conditions}
The HRT used a receiver stationed 15 km away from two transmitter units (at altitudes of $\sim$2500 m for the transmitters and $\sim$2000 m for the receiver). Each transmitter consisted of a laser diode (638~nm), optical diffuser, and driver electronics. The lasers were driven in a 64-of-128 pattern ID with a $\tau=\SI{2}{\micro\second}$ pulse duration, a $T=\SI{500}{\micro\second}$ pulse interval, which repeats the ID every \SI{64}{\milli\second} (recall Figure \ref{figwaveform}). The two transmitter units encoded different IDs. The time-averaged effective isotropic radiated power (EIRP) in the direction of the receiver was measured to be 0.3 mW, corresponding to a peak EIRP of 150 mW after adjusting for the 1:500 duty fraction. Unit 2 was attenuated by a 13\%-transmission neutral density filter, while Unit 1 was left unfiltered.

The receiver consisted of a LANL-developed photon-counting camera \cite{Priedhorsky2005,Thompson2013} with an f = 400 mm lens and adjustable aperture (up to f/2.8 = 143 mm diameter). The camera has a quantum efficiency of about 3.9\% at 638~nm and sub-nanosecond timing resolution. The receiver was equipped with a 10-nm bandpass filter centered on the transmitter laser wavelength (638 nm) to reduce background. 
The observations were made on a moonless night under variable clouds, beginning about 90 minutes after sunset.  
Observations were interrupted by rain and fog at times, but the data discussed here was taken while conditions were clear with no obvious obscuration.  The optical path was within Los Alamos County, NM, USA (population 18,000) and light pollution was minimal.

\subsubsection{Link budget verification}
Link budget calculations (see Appendix B) predict that a $P_{ave}=\SI{2}{\milli\watt}$ ELROI beacon on a sunlit CubeSat from LEO at 1000-km range from the Fenton Hill Observatory 14" telescope and LANL sensor would produce 3.3 signal counts and 91 background counts per second. Applied to the horizontal range test, the same calculation predicts that Unit 1 at f/8 will produce 4.7 counts per second, and the attenuated Unit 2 will produce the same count rate at f/2.8. As discussed in Appendix B, the variable transmission of the atmosphere is not accounted for in these estimates; however, the HRT test range of 15 km was chosen to provide more than the equivalent atmospheric attenuation of a sea-level site observing at the zenith.

The measured values from the HRT are 8.3 counts/s for Unit 1 at f/8 and 4.5 counts/s for Unit 2 at f/2.8.  Measurements covered a range of apertures yielding count rates from about 0.1 counts/s (attenuated Unit 2 at f/22) to 40 counts/s  (Unit 1 at f/2.8). The measured count rates for Unit 2 agree well with predictions, while count rates for Unit 1 are somewhat higher than predicted. This may be due to the effect of indirect light from the sides of Unit 1 illuminating the area around the transmitter and then reaching the receiver. In contrast, Unit 2 was housed in a box with a window covered by the neutral density filter. 

\subsubsection{Signal extraction and error rates}

To decode the ID, the photons from the location of the transmitter are first selected from the camera FOV. In the analysis of the HRT this is done manually with a radius cut around each transmitter location. As discussed in Appendix A, an algorithm is used to detect the carrier frequency of 2 kHz (see Figure \ref{figcarrier}), and out-of-phase photons are rejected. The remaining photon detections are then stacked by the 128-bit pattern length to form a histogram of counts in the cyclic time bins corresponding to each bit. A threshold level is set to distinguish between 0 and 1 values. Due to errors from statistical fluctuations and background, this analysis will give some fraction of incorrect bit values (the bit error ratio, or BER), so identifying the correct ID requires searching the registry for the number that has the fewest discrepancies from the measured sequence. For the HRT, an ID is considered correct if it contains 12 or fewer bit errors--an ID with 13 or more bit errors is considered a Codeword Error.  Optimal coding schemes can correct more than 12 erroneous bits out of 128, and using a simple threshold cut decreases statistical power compared to more sophisticated methods, therefore this is a conservative criterion.

Figure \ref{figcoderror} shows the bit error ratio (BER) as a function of observation time for the HRT conditions, and scaled by signal count rate to the predicted LEO system. The codeword error ratio (CER) is a function of the BER; as a result of using error-correcting codes, a BER of 3.7\% gives a CER of just one in a thousand while a BER of 1\% gives a CER below one in a billion. In the data of Figure \ref{figcoderror}, 55-95 seconds at the predicted LEO rate is required to achieve CER~$ = 10^{-3}$, and 105-157 seconds reduces the CER to $10^{-9}$. These observation times are reasonable for a single pass of a LEO satellite; therefore, we expect to be able to read the ELROI ID with high confidence in the predicted LEO system. The measured count rates in the HRT conditions also confirm that the link budget calculation methods are accurate. 
The HRT therefore successfully demonstrates the ELROI concept and lays the groundwork for the first on-orbit test, ELROI-PC104, described in the next section.

\subsection{ELROI-PC104}\label{pc104}

ELROI has been developed in a PC-104 form factor for integration into a CubeSat (Figure \ref{fig:prototypes}a).
It has 4 laser diode emitters projecting through two diffusers on opposite sides of the CubeSat.
One diffuser has two $\lambda=$ \SI{638}{\nano\meter} red laser diodes, at $P_{peak} = $ \SI{1}{\watt} and \SI{0.7}{\watt}, from two different manufacturers (Mitsubishi ML501P73 and Oclaro HL63193MG, respectively).
The other diffuser has an identical \SI{1}{\watt}  $\lambda = $ \SI{638}{\nano\meter} laser diode and a  $\lambda = $ \SI{450}{\nano\meter}  \SI{1.6}{\watt} blue laser diode (Thorlabs L450P1600MM).

This unit relies on the spacecraft bus for power, and can be controlled by the spacecraft computer over an Inter-Integrated Circuit (I$^{2}$C) interface.  Each laser emitter can be independently controlled with its own combination of pulse width $\tau$, pulse interval $T$, and ID number (up to 128 bits).  The relative phases of the diodes can be controlled so that they do not flash simultaneously, even if they have the same $T$.

In the absence of any commands from the spacecraft, the unit waits for 45 minutes after power is applied, then automatically powers up the three red laser diodes, each with its own ID and phase.  The 45 minute delay before transmission is required for 1-3U CubeSats by the CubeSat standard \cite{CalPoly2009}.  Only the red diodes are automatically activated because the blue laser requires a higher operating voltage and more power from the spacecraft bus.  In this three-diode, red-only mode, the entire ELROI unit draws only \SI{56}{\milli\watt}.

Many CubeSats fail after launch before providing their first telemetry \cite{Swartwout2013}.
Although the failure rate is declining as the community gains greater experience, the lack of feedback can prevent anomalies from being diagnosed, making recovery or learning from the experience very difficult.
The autonomous start-up of ELROI-PC104 will improve this situation.
If the ELROI signal is detected, it will demonstrate that the spacecraft power systems are active.
As the spacecraft processor boots up and starts activating systems, it can update the transmitted ID, allowing progress to be monitored.
Any failures detected by the processor can also be transmitted by modifying the ELROI ID.
Other status information, such as the number of radio commands received from the ground, can be transmitted in the same way.

The unit shown in Fig \ref{fig:prototypes}a has been integrated into the New Mexico Institute of Technology's NMTSat, a 3U CubeSat with launch expected in 2018 \cite{nmtsat}.

\subsection{ELROI-UP}\label{elroiup}

A unit that can be attached to any satellite, the ELROI Universal Prototype (ELROI-UP), is currently being designed and built (Figure \ref{fig:prototypes}b).  
This model includes its own solar cell and battery, and is capable of fully autonomous operation.  
It can accept commands or power from the host spacecraft, but does not require them.  
The unit can also be mounted in a passive mechanical structure and launched as a free-flying CubeSat in 1/3U, 1/2U or larger form factor.

ELROI-UP supports up to four laser diodes.  
The first test flight units will be populated with four $\lambda = $ \SI{638}{\nano\meter} red laser diodes, at $P_{peak} = $ \SI{2.5}{\watt} (Mitsubishi ML562G84).
Multiple codes and timings with different combinations of the emitters will be pre-programmed to allow testing at up to $P_{peak} = $ \SI{10}{\watt}.

The first production run will be of 6 units at a marginal cost of \$4,000 each.  The size is $9.8\times9.2\times3.1 = \SI{280}{\centi\meter^3}$, the mass is \SI{300}{g} and the power (if externally supplied instead of provided by the solar cell) is \SI{50}{\milli\watt}.  This \$SWAP qualifies it as a Low Resource Optical Identifier.  We are willing to provide these beacons to appropriate launch opportunities.

\subsection{ELROI-1.0}\label{elroi1}

Test results from ELROI-PC104 and ELROI-UP will allow ELROI to be standardized and adopted by the entire space community.  CubeSats are the class of space object that have the most pressing need for an identification system, and are also the most constrained by \$SWaP.  Thus we need to develop the lightest, smallest, cheapest ELROI beacon that can be mass-produced and attached to any CubeSat with minimal integration cost.  

The size of beacon will be dominated by its largest components, which in this case will be the power system.  A given system power mandates the solar cell area needed to supply it, and the battery volume required to maintain power both through the orbital night, and through extended periods when a randomly-oriented inert or non-attitude-controlled satellite might keep the solar cell pointed away from the Sun.  A satellite that maintains attitudes that prevent good solar illumination (\emph{e.g.}, if the beacon is on a nadir-pointing face of the spacecraft) may require additional, separated, or re-oriented solar cells.

For the LEO CubeSat beacon used as an example in the link budget Appendix B, $P_{ave}$ is \SI{2}{\milli\watt} of optical power.  
Assuming a 50\% conversion efficiency from electricity to light and allocating an additional \SI{1}{\milli\watt} for other components such as the pulse pattern generator, this gives a total system power of \SI{5}{\milli\watt}.  
We assume a photovoltaic cell that is 30\% efficient, is randomly oriented on the relevant timescales (50\% of the time it is pointed away from the Sun, and the remaining time has a slant-adjusted area that averages to 50\%) and is in eclipse for 50\% of its orbit.
These benchmark values combined with a \SI{1.36}{\kilo\watt\per\m^2} solar constant give an average yield of \SI{5}{\milli\watt\per\cm^2} of photovoltaic cell area.
LiFePO4 cells store $\sim$\SI{0.22}{\watt\hour\per\cm^3} of power (e.g., \cite{lifepo4}) so \SI{0.55}{\centi\meter^3} of this chemistry could power the beacon for an entire day on a full charge. 
Thus, if the size of the beacon is dominated by the power system, this implies a size scale of $\sim$\SI{1}{\cm^3}.  

Although this minimal-size system would provide adequate power and signal under the typical conditions, increasing the overall signal would allow the ID to be read over a wider range of conditions, such as shorter-duration passes at lower horizon angles and longer distances from the ground station.  Under optimal conditions, an increased signal would allow the ground station to receive the ID code more rapidly, allowing it to service more satellites per night.   Allowing a larger beacon would also simplify the engineering challenges of manufacture.   Experience gained from the early flights of ELROI-PC104 and ELROI-UP will help to refine the practical trade-offs between the cost and benefits of a larger, more powerful beacon.

Figure \ref{fig:prototypes}c shows a mock-up of what such a beacon could look like in a $2 \times 2 \times 0.5 ~\si{\cm^3}$ size with $\SI{3}{\cm^3}$ of photovoltaic cell.  Attaching one or more of these to every CubeSat would be technically and economically feasible and, given the benefits that it provides to the spacecraft operator, likely to be widely adopted.

\section{Conclusion}
\label{sect:conclusion}

ELROI is a simple, low-cost optical beacon that allows a modest ground station to provide unambiguous Space Object Identification.  The beacon is cheap and small enough for use on CubeSats, and ground stations can be built around the world using commercially-available technology.  It is suitable for use on everything that goes into space to simplify the important task of Space Traffic Control.  

The ELROI concept has been verified in long-range ground tests that simulate the link budget and atmospheric path-length of a LEO system.  A beacon has been integrated into a spacecraft for launch in the near future, and will be observed with an existing ground station.  Multiple copies of an autonomous beacon that can be attached to other spacecraft with minimal integration are being built.  There are no technical barriers to miniaturizing ELROI to the size of a thick postage stamp for easy application to even small satellites.

We encourage other groups to develop their own ground systems or use existing facilities to track the ELROI beacons after launch.  We can also provide prototype beacons to spacecraft developers for integration into their satellites.

Based on the data from these initial flights, international standards can be developed to allow any ground station to identify an unknown satellite that has a beacon.  Producing an optimal set of requirements on the beacon and ground stations will also require extensive physical modeling and simulation, vendor surveys, technology forecasts, economic and policy consideration, and value judgements to balance the competing interests that will eventually arise. 

The advantages that ELROI provides to spacecraft operators, combined with the low expenditure required to incorporate the beacons in satellites, means that they can become widely used even in the absence of international norms or mandates.

\section*{Appendix}
\subsection{Data extraction and analysis}
\label{subsect:datareading}

This is a simplified description of the process of determining an ELROI beacon's ID from observations.
A more detailed discussion of statistical analysis and computational optimization techniques is beyond the scope of this paper.

The telescope and detector acquire the data from the beacon as a list of time-tagged photon detection events.
To go from that data to a ID is a multi-step process: 
\begin{enumerate}
  \item Detect and recover clock phase and period.
  \item Extract photon count values for each ID bit.
  \item Determine the ID that best fits the data, and an estimate of its reliability.
\end{enumerate}

If the precise clock period of the beacon, $T$, is known, then the fractional phase of each photon in the range of $[0,1)$ cycles, is
$$\phi_{i} = \textrm{frac}(t^{\prime}_{i} / T)$$
where $t^{\prime}_{i}$ is the time at which photon $i$ is detected, adjusted for light-travel time from the satellite's location, and 
$\textrm{frac}(x) = x - \lfloor {x} \rfloor$ is the fractional part function.
The true clock phase may be determined by generating a histogram of $\phi_{i}$ and finding the maximum at a value $\phi_{\textrm{peak}}$, with a width $\tau / T$.

In practice, even if a nominal value of $T_{\textrm{nom}}$ is standardized (so that it can be predicted for an unidentified satellite), the true value of $T$ will be in an uncertainty range $T_{\textrm{nom}} (1-e_T) < T < T_{\textrm{nom}} (1+e_T)$ where $e_T$, the fractional error tolerance in $T$, is due to tolerances in the beacon's clock.
For a crystal oscillator, $e_T \sim 50 \times 10^{-6}$ or 50 ppm is a common value.
In addition, there may be drift in $T$ during the course of the observation, primarily due to temperature variation.

This variation in $T$ can be handled by searching over the expected range for the value that produces the best peak in the histogram of $\phi_{i}$.
To prevent loss of sensitivity, this search will require a granularity of $\delta T \lesssim \tau T / { \Delta t}$ where $\Delta t$ is the total duration of the data analyzed.
For a \SI{10}{\second} observation with $\tau =$ \SI{1}{\micro\second}, $T_{\textrm{nom}} =$ \SI{1}{\milli\second}, a $\delta T = 1 \times 10^{-10} s$ search over a $\pm 50$ ppm range ($T_{\textrm{nom}} \pm 5 \times 10^{-8}$) will require 1000 trial histograms.

This can be handled with good computational efficiency by techniques such as the Fast Folding Algorithm \cite{Staelin1969} (FFA), which allows a specified frequency range to be searched with a specified frequency spacing to look for features that are narrow in phase space.  The FFA can be extended to allow efficient searches over slow changes in frequency during the observation to compensate for temperature drift.  The FFA is fast enough that detecting and recovering the clock phase will take an insignificant amount of computer time compared to the duration of the observation.

When a trial histogram shows a significant peak, small adjustments to $T$, drift in $T$, and the phase offset can often improve the peak value if the beacon photons were originally distributed among adjacent phase bins.  This `peaking up' is an acceptable way of refining the clock solution, with the understanding that the significance of the result reflects a larger number of trials.

Note that the basic version of the much more familiar Fast Fourier Transform (FFT) is not computationally efficient for this search.  
This is because the spectral energy of the ELROI signal is spread out over many harmonics (from below a \SI{}{\kilo\hertz} to above a \SI{}{\mega\hertz}), 
the FFT must be calculated over all frequencies from zero to the highest frequency of interest, the FFT's frequency spacing is fixed and inadequate, and the FFT cannot easily handle slowly-varying frequencies.

After the clock characteristics are determined and refined, the photon detections can be divided into those whose detection times are respectively in-phase and out-of-phase with the beacon's emission.  The out-of-phase photons are used to determine the background rate, but can be otherwise discarded. 
 The in-phase photons, which include both beacon and background photons are used in further analysis.  
 Each in-phase photon detection is given a bit index 
 $0 \leq j^{\prime} < m $ 
 which indicates which bit it corresponds to in the repeating 
 $m$-bit ID, with the prime ($^{\prime}$) to indicate that the 'first' bit of the ID as found in the registry does not necessarily correspond to $j^{\prime} = 0$.
 
$C_{j^\prime}$, the number of detected photons for bit index $j^{\prime}$, is therefore a measurement of whether the corresponding bit of the ID is a one or a zero.
This will tend to be bimodally distributed, with the expectation value for bits that are zero, $\langle^{0}N\rangle$, being estimable from the background rate, and the expectation value for bits that are one having an additional signal: 
$\langle^{1}N\rangle = \langle^{0}N\rangle + \langle^{1}S\rangle$.

The simplest way to match the data to an ID is to assign a threshold cut:
\[
 b_{j^\prime} = 
 \begin{cases}
	 1, & \text{if } C_{j^\prime} \geq C_{\textrm{thresh}}\\
	 0, & \text{otherwise}
 \end{cases}
\]
with $C_{thresh}$ chosen to split the $b_{j^\prime}$ values into the expected number of zeros and ones.

The set of  $b_{j^\prime}$ values is then matched to each individual IDs in the catalog of all currently flying ELROI units, with each of the $m$ bit positions being used as $j^{\prime} = 0$.  The ID that matches the set of $b_{j^\prime}$ with the fewest discrepancies is the most likely ID to correspond to the observed ELROI beacon.  If all other IDs have substantially more discrepancies, then that indicates with high confidence that the identification is correct.

There are many possible refinements to this algorithm.  Improved clock recovery algorithms that take advantage of the sparse nature of the data (the relatively small number of detected photons) can be developed.  Photon detections can be weighted based on the measured variation in the background and expected signal level.  The threshold cut may be replaced by a probabilistic analysis with a continuous rather than binary 0 vs 1 determination.  The confidence level of an identification can be rigorously calculated based on the counting statistics of the photon detections.  These refinements significantly improve the beacon detectability and are straightforward to implement, but are beyond the scope of this paper.

\subsection{Link budget calculation for a sample design}
\label{subsect:linkbudget}

The link budget gives the rate of signal photon detections (\si{photons\per\second}) expected under various conditions. This rate depends on the optical power emitted by the beacon, the distance from the beacon to the ground station receiver, the area of the collecting optics, the transmission of the atmosphere and optical components, and the quantum efficiency of the photon-counting sensor. The values in the following example calculation are based on the LANL ground station at Fenton Hill, but would be applicable to most mid-sized telescopes and many different types of single-photon detectors. An estimate of background count rates in several scenarios is also given.

The count rate in \si{photons\per\second} at the receiver is
$$
R = P_{\textrm{avg}} \times 1/\Omega \times A/r^2 \times T_{\textrm{tot}} \times \varepsilon_{\textrm{DQE}}/E_{\gamma}
$$
$P_{\textrm{avg}}$ is the average power of the beacon, which depends on the peak power $P_{\textrm{peak}}$ and the duty cycle $d$. $\Omega$ is the solid angle of the beacon emission, and can be as large as 2$\pi$. $A$ is the collecting area of the receiver optics, and $r$ is the distance from the beacon to the receiver. $T_{\textrm{tot}}$ is the total optical transmission, including the spectral filter transmission $T_{\textrm{f}}$ and the atmosphere transmission $T_{\textrm{atm}}$. Atmospheric transmission can change rapidly and depends on so many factors (atmospheric density, humidity, pollution, zenith angle, wavelength, etc.) that a detailed model is beyond the scope of this work. Thus, as in \S \ref{hrt}, $T_{\textrm{atm}}$ is ignored for the purposes of a link budget estimate, but it can be expected to contribute 1-3 dB of additional loss depending on conditions. $\varepsilon_{\textrm{DQE}}$ is the quantum efficiency of the photon-counting detector, which ranges from 1\% (some PMTs) to 75\% (some SPADs), and has been measured to be 3.9\% for the LANL sensor. $E_{\gamma}$ is the energy per photon at the beacon wavelength.

Typical values for the link budget parameters are shown in Table \ref{table:link-budget-parameters}, and the calculation is worked out in Table \ref{table:link-budget-calc}, giving a nominal rate of 3.3 \si{photons\per\second} from LEO.

\input{parameters-table.tex}

\input{beacon-table.tex}

When the host satellite is in sunlight, reflected light will be the dominant source of background photons \footnotemark .
The amount of reflected sunlight depends on a combination of the satellite's size and albedo, and will generally be higher for larger satellites.  This background also depends on the spacecraft attitude and the angle between the Sun and the observer, so for simplicity the spacecraft is approximated by an equivalent sphere at a 90\textdegree~(``half moon'') phase angle.  The background rate for a 10-\si{\centi\meter} sunlit CubeSat is worked out in Table \ref{table:background}, giving a nominal rate of 0.36 background photons/second after filtering.

\footnotetext{The contribution from sky background is usually small, but depends on the local sky surface brightness and the field of view necessary for tracking. Assuming a field of view of 1 arcminute and the other parameters given in Table \ref{table:link-budget-parameters}, the sky background count rate ranges from about 40 \si{photons\per\second} with no moonlight (sky surface brightness = 21.6 \si{magnitude\per}arcsecond$^2$) to about 1000 \si{photon\per\second} at the full moon (sky surface brightness = 18 \si{magnitude\per}arcsecond$^2$), before the phase cut \cite{Spoelstra2009}. (Note that sky surface brightness values are measured in the V spectral band; scattering of light in the atmosphere is biased toward the blue end of the visible spectrum, so count rates at 638 nm will be lower.) After the phase cut, these rates would be reduced to about 0.2 \si{photons\per\second} and 4 \si{photons\per\second}, respectively. The background count rate may increase dramatically if bright stars enter the field of view, but these events are isolated in time and can be removed from the data.}

\input{background-table.tex}

The predicted signal rate of 3.3 \si{photons\per\second} and background rate of 0.36 \si{photons\per\second} gives a sufficient link budget to read the ID reliably in a single pass---as discussed in \S\ref{hrt} and shown in Figure \ref{figcoderror}, 55 to 95 \si{\second} of observation at the predicted LEO rate is required to reduce the CER (Codeword Error Ratio, the probability of incorrectly identifying the code) to one in a thousand, and 105 to 157 \si{\second} reduces the CER to one in a billion.

\section*{Acknowledgments}

Initial work on this project was supported by the US Department of Energy through the Los Alamos National Laboratory (LANL) Laboratory Directed Research and Development program as part of the IMPACT (Integrated Modeling of Perturbations in Atmospheres for Conjunction Tracking) project.  Further work was supported by the Richard P. Feynman Center for Innovation, LANL.  ELROI hardware and software was developed at LANL by Louis Borges, Richard Dutch, David Hemsing, Joellen Lansford and Charles Weaver, with thermal analysis by Alexandra Hickey, Lee Holguin, and Zachary Kennison. The Horizontal Range Test was assisted by Amanda Graff, David Graff, Mike Rabin, and David Thompson.  We thank the New Mexico Tech CubeSat team, including Anders Jorgensen, Sawyer Gill, Samantha Young, and Aaron Zucherman, for providing the first opportunity to fly an ELROI unit.  We thank Roberta Ewart of the Space and Missiles Systems Center for her encouragement and advocacy.

\section*{References}
\bibliography{ref}
\bibliographystyle{aiaa}

\end{document}

%% file: parameters-table.tex
\begin{table}[p!]
\centering
\caption{Example link budget and background parameters for the ELROI beacon. The effective albedo areas of sunlit host satellites assume the object is a diffuse white reflecting sphere with an albedo of 1, viewed at a 90-degree phase angle (``half moon''); actual albedo may be lower. \cite{Cognion2013,Vallerie1963}.}
\label{table:link-budget-parameters}
\begin{tabular}{llll}
\\ \hline
\textbf{Link budget parameter}               &                                 & \textbf{Value}       \\ \hline
Beacon wavelength                            & $\lambda$                       & 638 nm               \\
Peak power emitted                           & $P_{\textrm{peak}}$             & 1 W                  \\
Pulse width                                  & $\tau$                          & 2 $\mu$s             \\
Pulse interval                               & $T$                             & 500 $\mu$s           \\
Fraction of 1 bits                           & $f_1$                           & 0.50                 \\
Solid angle of emission                      & $\Omega$                        & $2\pi$               \\
Distance to receiver                         & $r$                             & 1000 km              \\
Diameter of receiver telescope               & $D$                             & 36 cm                \\
Filter transmission at $\lambda$             & $T_{\textrm{f}}$                & 0.83                 \\
Filter bandwidth                             & $\Delta\lambda$                 & 10 nm                \\
Solar spectral flux at $\lambda$             & $I_{\lambda}$                   & 1.654 W/m$^2$/nm     \\
Detector quantum efficiency                  & $\varepsilon_{\textrm{DQE}}$    & 0.039                \\
10-cm Cubesat effective albedo area          & $\alpha_{\textrm{CS}}$          & 0.00053 m$^2$        \\
1-m satellite effective albedo area          & $\alpha_{\textrm{1m}}$          & 0.053 m$^2$          \\
\end{tabular}
\end{table}

%% file: beacon-table.tex
\begin{table}[p!]
\centering
\caption{Example link budget calculation for the ELROI beacon. The expected count rate is the product of all the contributions given below. With the parameters given in Table \ref{table:link-budget-parameters}, the expected signal count rate is 3.3 photons/second, or 5.2 dB$\gamma$ (where the unit dB$\gamma$ is referenced to 1 photon/second).}
\label{table:link-budget-calc}
\begin{tabular}{l l l l r l}
\\ \hline
\textbf{Link budget contribution} & \textbf{Factor}                &                                     & \textbf{Value}                  & \textbf{dB}    & \textbf{Reference}     \\ \hline
Peak power                        & $P_{\textrm{peak}}$            &                                     & 1000 mW                         & 30             & 1 mW                   \\
Average power (duty cycle)        & $d$                            & = $\tau/T \times f_1$               & 0.002                           & -27            & ---                    \\
Solid angle of emission           & $1/\Omega$                     &                                     & $1/2\pi$                        & -8             & 1 sr$^{-1}$            \\
Distance to receiver              & $1/r^2$                        &                                     & $10^{-12}$ m$^{-2}$             & -120           & 1 m$^{-2}$             \\
Receiver aperture                 & $A$                            & = $\pi (D/2)^2$                     & 0.1 m$^2$                       & -10            & 1 m$^2$                \\
Filter transmission               & $T_{\textrm{f}}$               &                                     & 0.83                            & -0.8           & ---                    \\
Detector quantum efficiency       & $\varepsilon_{\textrm{DQE}}$   &                                     & 0.039                           & -14            & ---                    \\
Energy per photon                 & $1/E_{\gamma}$                 & = $(hc/\lambda)^{-1}$               & $3.2 \times 10^{15}$ photons/mJ & 155            & 1 photon/mJ            \\ \hline
\textbf{Expected count rate}      &                                &                                     & 3.3  photons/second             & 5.2            & 1 photon/s (dB$\gamma$)           
\end{tabular}
\end{table}

%% file: background-table.tex
\begin{table}[p!]
\caption{Example background rate calculation. For a sunlit 10-cm Cubesat host with effective albedo area $\alpha_{\textrm{CS}}$, the expected background count rate (shown below) is 0.36 photons/second. For a sunlit 1-m satellite with albedo area $\alpha_{\textrm{1m}}$, the expected background count rate is 36 photons/second. Solar spectral flux at $\lambda$ taken from ASTM standard \cite{ASTM}.}
\label{table:background}
\centering
\begin{tabular}{l l l l r l}
\\ \hline
\textbf{Background contribution}         &                              &                             & \textbf{Value}                  & \textbf{dB}      & \textbf{Reference} \\\hline
Solar spectral flux at $\lambda$         & $I_{\lambda}$                &                             & $1.654 \times 10^3$ mW/m$^2$/nm & 32.2             & 1 mW/m$^2$/nm      \\
Filter bandwidth                         & $\Delta\lambda$              &                             & 10 nm                           & 10               & 1 nm               \\
CubeSat effective albedo area            & $\alpha_{\textrm{CS}}$       &                             & 0.00053 m$^2$                   & -32.7            & 1 m$^2$            \\
Distance to receiver                     & $1/r^2$                      &                             & $10^{-12}$ m$^{-2}$             & -120             & 1 m$^{-2}$         \\
Receiver aperture                        & $A$                          & = $\pi (D/2)^2$             & 0.1 m$^2$                       & -10              & 1 m$^2$            \\
Transmission                             & $T_{\textrm{f}}$             &                             & 0.83                            & -0.8             & ---                \\
Detector quantum efficiency              & $\varepsilon_{\textrm{DQE}}$ &                             & 0.039                           & -14.1            & ---                \\
Energy per photon                        & $1/E_{\gamma}$               & = $(hc/\lambda)^{-1}$       & $3.2 \times 10^{15}$ photons/mJ & 155              & 1 photon/mJ        \\ \hline
\textbf{Expected background count rate}  &                              &                             & 91 photons/second               & 19.6             & 1 photon/s (dB$\gamma$)   \\ 
\textbf{before phase cut}                &                              &                             &                                 &                  &                    \\ \hline
Phase cut                                & $f_{\phi}$                   & $\tau/T$                    & 0.004                           & -24              & ---                \\ \hline
\textbf{Expected background count rate}  &                              &                             & 0.36 photons/second             & -4.4             & 1 photon/s (dB$\gamma$)        \\
\textbf{after phase cut}                 &                              &                             &                                 &                  &   
\end{tabular}
\end{table}